\documentclass[twocolumn,showpacs,amssymb]{revtex4}

\usepackage{graphicx}
\usepackage{psfig}
\usepackage{dcolumn}
\usepackage{bm}

\def\lessim{\lower.5ex\hbox{$\; \buildrel < \over \sim \;$}}
\def\gtrsim{\lower.5ex\hbox{$\; \buildrel > \over \sim \;$}}

\begin{document} 
\topmargin -0.8cm
\preprint{}

\title{Limit on Quark-Antiquark Mass Difference from the Neutral Kaon System}

\author{Michael J. Fromerth and Johann Rafelski}
\affiliation{Department of Physics, University of Arizona, Tucson, AZ 85721}
\date{April 2003}

\begin{abstract}
We quantify the limits on quark-antiquark mass differences 
imposed by the neutral kaon mass system.
In particular, we find that an upper limit to the 
mass difference of $10^{-3}$\,eV exists if mass 
differences across quark flavors are uncorrelated.
In the upcoming antihydrogen 
experiments this limit on quark mass difference would allow 
a measurement of electron-positron mass 
difference up to a relative precision level of $10^{-15}$.
\end{abstract}

\maketitle

The origin of quark and lepton masses remains at present 
unknown. It is generally presumed that by virtue of CPT
symmetry matter and antimatter will always have same fundamental
mass parameters. On the other hand, 
the experimental limits are at present 
not very good, and in fact the limit on the  baryon-antibaryon 
mass difference~\cite{PDG02}:
$$|m_p-m_{\bar p}|< 6\,10^{-8}m_p\simeq 60\,\mbox{eV}.$$
is much grater than what would suffice to induce a small
baryon-antibaryon abundance 
asymmetry in the early Universe evolving into
present time. At the deconfinement boundary a 
baryochemical potential of the order of
 $1.1\mbox{eV}\pm15\%$~\cite{From02} suffices.
Consequently, it would seem that our understanding 
of the early Universe depends
sensitively on the implicit assumption about the presence of 
the mass symmetry for quarks and antiquarks, the constituents of 
baryons, at a level of two orders of magnitude beyond current
experimental knowledge.  

Here we show how the properties of the neutral kaon system 
constrain much more accurately
the mass difference between quarks and antiquarks.
We also comment on achievable  improvement in the 
measurement of matter-antimatter symmetry in
comparisons of hydrogen with antihydrogen.

From a detailed study of the kaon decay rates, it is observed 
that the mass difference between the $K_L$ and $K_S$ states 
is $\Delta m \equiv m_{K_L} - m_{K_S} 
= 3.463 \pm 0.010 \times 10^{-6}$\,eV~\cite{Alavi02}.
Because this mass difference is understood within the standard 
model to arise from second-order weak interactions that mix 
the $K^0$ and $\bar{K^0}$ states~\cite{Gaillard75,Perkins00}, the 
magnitude of the CPT-violating contribution to
particle-antiparticle mass difference is severely 
restricted by this result.

Recently, it has been demonstrated by Greenberg~\cite{Greenberg02} 
that CPT breaking implies the violation of coordinate Lorentz invariance. 
In an extension of the standard model, Lorentz- and CPT-violating 
operators yielding a satisfactory quantum field theory have been
considered \cite{Colladay97,Colladay98,Kostelecky01a}. 
Examples include spontaneous Lorentz and CPT violation in the 
context of string field theory (see, e.g., Refs.~\cite{Kostelecky89,
Kostelecky91}) and non-commutative field theories~\cite{Carroll01}.
These models offer a basis for numerous precision experiments 
placing extremely tight bounds on Lorentz and CPT breaking. 
In this context, the neutral kaon system has been analyzed both
experimentally~\cite{Carosi90,Schwingenheuer95} and 
theoretically~\cite{Kostelecky98,Kostelecky99,Kostelecky01b}, while 
direct CPT violation in the neutrino sector has been explored in 
Refs.~\cite{Skadhauge02,Barenboim02a,Barenboim02b}.
Other work discussing theoretical implications of CPT violation 
include~\cite{Barmin84,Dejardin00,Klinkhamer01,Urbanowski02}.

One may indeed ask if such a hypothesis for a mass difference
between particles and antiparticles 
 makes good physical sense, considering the well 
established principles of quantum field theory.
We believe that  a search for tacitly assumed limits to accepted
physical principles  is a very important step in 
verification of the paradigm which govern our view on the 
laws of physics. This attitude is more generally shared, and
with formation of a large number of antihydrogen atoms, we
can look forward to further experimental  tests of CPT symmetry  to 
take place at CERN~\cite{CERN02a,CERN02a}. On a more 
theoretical side, a momentum-dependent difference 
 between particles and antiparticles is 
expected~\cite{Kostelecky98,Kostelecky99}, should there be 
a violation of Lorentz invariance, appearing in association 
with  CPT breaking~\cite{Greenberg02}. 

In the following we describe 
how the measured mass asymmetry of the  neutral kaons limits
the mass difference between quarks and antiquarks.
We define the $K_L$ and $K_S$ states in the standard formalism~\cite{Perkins00}:
\begin{eqnarray}
K_L & = & \frac{1}{\sqrt{2 + 2 \epsilon^2}} 
  \left[ (1+\epsilon) K^0 + (1-\epsilon) \bar{K^0} \right] \ , \label{K_L} \\
K_S & = & \frac{1}{\sqrt{2 + 2 \epsilon^2}} 
  \left[ (1-\epsilon) K^0 - (1+\epsilon) \bar{K^0} \right] \ ,
\end{eqnarray}
where $K^0 = |d\bar{s}\rangle$, $\bar{K^0} = |\bar{d}s\rangle$, and 
$\epsilon \approx 2.3 \times 10^{-3}$ is the CP violation parameter.

We express the (assumed) CPT-violating mass difference between quarks and antiquarks as:
\begin{eqnarray}
m_{s,\bar{s}} & = & m_s^0 \pm \frac{\delta m_s}{2} \label{delta_ms} \\
m_{d,\bar{d}} & = & m_d^0 \pm \frac{\delta m_d}{2} \label{delta_md} \ ,
\end{eqnarray}
where the signs of $\delta m_s$ and $\delta m_d$ are undetermined.

The mass operator for the neutral kaon system, with the quark mass differences, becomes:
\begin{eqnarray}
\hat{M}\ =\ \hat{M}^0_{K}\ +\ \hat{M}_{w}\ +\ \frac{f}{2}\, 
 \left[\, (\delta m_d - \delta m_{\bar{s}})\, |d\bar{s}\rangle 
\langle d \bar{s}|\ +\ \right.\nonumber\\ \left.
+\ (-\delta m_{\bar{d}} + \delta m_s)\, |\bar{d} s \rangle 
\langle \bar{d} s|\, \right] ,\ \
\label{mass}
\end{eqnarray}
where $\hat{M}^0_{K}$ is the neutral kaon mass excluding weak interactions, 
$\hat{M}_{w}$ is the mass contribution due to weak interactions, and the 
third term is the effect that the change in the current quark masses of 
Eqs.~(\ref{delta_ms}) and (\ref{delta_md}) would have on the kaon mass.

The form of the third term arises because in a model of hadronic structure 
(e.g., the bag model), the response of the hadronic mass is linear with 
respect to the change in quark mass if expanded about 
a finite quark mass~\cite{Letessier02}.
Note that a similar effect arises in non-relativistic quark models.
Furthermore, the scaling factor $f$ is of order unity, as confirmed by 
the features of hadronic mass splittings.

  From Eqs.~(\ref{K_L})--(\ref{mass}), 
the mass difference between $K_L$ and $K_S$ becomes:
\begin{eqnarray}
\Delta m & = & \langle K_L | \hat{M} 
  | K_L \rangle \ -\ \langle K_S | \hat{M} | K_S \rangle \nonumber \\
 & = & \Delta m_{w}\ +\ 2\, \epsilon f\, 
  \left[\, (m_{\bar{s}} - m_s) - (m_{\bar{d}} - m_d)\, \right]
\end{eqnarray}
where 
$\Delta m_{w} \equiv\langle K_L | \hat{M}_w 
  | K_L \rangle - \langle K_S | \hat{M}_w | K_S 
\rangle$ 
and terms of $\epsilon^2$ or higher have been neglected.
Since it is understood that $\Delta m \simeq \Delta m_{w}$, this immediately 
yields the result:
\begin{equation}
\left| (m_{\bar{s}} - m_s) - (m_{\bar{d}} - m_d) \right| \ \ll\ 
\frac{\Delta m}{2 \epsilon f}\ \approx\ 10^{-3}\, {\rm eV} \ .
\label{direct_cpt}
\end{equation}

Equation~(\ref{direct_cpt}) places rather stringent limits 
on direct CPT violation in $d$ and $s$ quarks.
If the size of the CPT violation across quark flavors is 
uncorrelated, then an upper limit to the mass difference 
between quarks and antiquarks of {\it each} flavor must be much 
less than $10^{-3}$\,eV.
Otherwise, the size of the CPT violation across $s$ and $d$ 
flavors must be highly correlated, such that 
$(m_{\bar{s}} - m_s) \simeq (m_{\bar{d}} - m_d)$. 
This would imply that a CPT violating force does not in effect
distinguish between the first and second particle generation.
In the following, we assume that this is not the case.

Such a small mass difference between $d$ and $\bar{d}$ 
quarks allows a thorough study of the possible electron-positron mass difference
in the antihydrogen experiments.
The wavelengths of atomic transitions in hydrogen scale with the inverse 
of the reduced mass, $\lambda \propto (m_p + m_e)/m_e m_p$.
As a result, the relative shift in wavelength due to a mass difference 
in hydrogen and antihydrogen atoms comprises also, at a lesser degree, 
the influence of the atomic nucleus:
\begin{eqnarray}
\left| \frac{\delta \lambda}{\lambda} \right| 
& = & \frac{m_p m_e}{m_p + m_e} 
\left(\frac{\delta m_e}{m_e^2} + \frac{\delta m_p}{m_p^2} \right) \nonumber \\
 & \simeq & \left[\,\frac{1}{m_e}\, \delta m_e \ +\ 
\frac{m_e}{m_p^2}\, f\, \left(2 \delta m_u + \delta m_d \right)\, \right] \ . 
\label{waveshift}
\end{eqnarray}
A CPT violation  originating in  quarks and antiquarks is thus greatly 
reduced. Indeed, if the mass difference between $u$ and $\bar{u}$ is also
limited to be $\ll 10^{-3}$\,eV, then the resulting contribution of the atomic
nucleus to the shift in wavelength becomes:
\begin{equation}
\left| \frac{\delta \lambda}{\lambda} \right| \ll  2 \times 10^{-15} \ .
\end{equation}
This therefore is in principle the precision with which the 
relative mass difference of electron and positron can 
be measured in experiments involving matter and 
antimatter \cite{CERN02a,CERN02b}. Only when this precision is indeed
reached  we would also become sensitive in these experiments 
to the possible quark-antiquark mass differences. 

In summary, we find that the current upper limit to the mass difference 
between quarks and antiquarks in the $d$ and $s$ flavors is 
$\ll 10^{-3}\, {\rm eV}$ if the magnitude of the CPT 
violation is uncorrelated across quark flavors. 
In this case, the relative precision 
with which the strange quark mass difference is determined appears to be 
by far the most precise such value presently known:
\[
\left|\frac{m_s-m_{\bar s}}{m_s+m_{\bar s}}\right|\ll 10^{-11}\,,
\]
providing a strong constraint for any CPT model considered, and assuring
that a possible quark mass asymmetry is not relevant in the determination of
the physical conditions in the early Universe.

A possible $d$-quark mass difference at this level would have the effect 
of shifting the wavelengths of the antihydrogen atomic 
spectrum by $\ll  2 \times 10^{-15}$ relative to the hydrogen spectrum, allowing
a measurement of the mass difference in the leptonic 
sector at a yet much higher precision.

{\vspace{0.5cm}\noindent\it Acknowledgments:\\}
The authors wish to thank Elliott Cheu and Ralf Lehnert for their helpful comments.
Supported  by a grant from the U.S. Department of Energy,  DE-FG03-95ER40937\,. 


\begin{thebibliography}{10}

\bibitem{PDG02}
K.~Hagiwara, {\it et al.},
\newblock{Review of Particle Physics 2002},
\newblock {\em Phys. Rev.}, D66:010001, 2002.


\bibitem{From02}
M.J. Fromerth, and J. Rafelski
\newblock {Hadronization of the Quark Universe}
\newblock {\em astro-ph/0211346} 



\bibitem{Alavi02}
A.~{Alavi-Harati} et~al.
\newblock {Measurements of Direct CP Violation, CPT Symmetry, and Other
  Parameters in the Neutral Kaon System}.
\newblock {\em  Phys.Rev.}, D67:01200501--01200533, 2003.

\bibitem{Gaillard75}
M.~K. {Gaillard}, B.~W. {Lee}, and J.~L. {Rosner}.
\newblock {Search for charm}.
\newblock {\em Rev. Mod. Phys.}, 47:277--310, April 1975.

\bibitem{Perkins00}
D.~H. {Perkins}.
\newblock {\em {Introduction to High Energy Physics}}.
\newblock Cambridge, 2000.

\bibitem{Greenberg02}
O.~W. {Greenberg}.
\newblock CPT violation implies violation of Lorentz invariance.
\newblock {\em Phys. Rev. Lett.}, 89:2316021--2316024, 2002.

\bibitem{Colladay97}
D.~{Colladay} and V.~A. {Kostelecky}.
\newblock CPT violation and the standard model.
\newblock {\em Phys. Rev.}, D55:6760--6774, 1997.

\bibitem{Colladay98}
D.~{Colladay} and V.~A. {Kosteleck{\' y}}.
\newblock {Lorentz-violating extension of the standard model}.
\newblock {\em Phys. Rev. D}, 58:116002--116024, December 1998.

\bibitem{Kostelecky01a}
V.~A. {Kosteleck{\' y}} and R.~{Lehnert}.
\newblock {Stability, causality, and Lorentz and CPT violation}.
\newblock {\em Phys. Rev. D}, 63:65008--65026, March 2001.

\bibitem{Kostelecky89}
V.~A. {Kosteleck{\' y}} and S.~{Samuel}.
\newblock {Phenomenological gravitational constraints on strings and
  higher-dimensional theories}.
\newblock {\em Phys. Rev. Lett.}, 63:224--227, July 1989.

\bibitem{Kostelecky91}
V.~Alan Kostelecky and Robertus Potting.
\newblock CPT and strings.
\newblock {\em Nucl. Phys.}, B359:545--570, 1991.

\bibitem{Carroll01}
S.~M. {Carroll} et~al.
\newblock Noncommutative field theory and Lorentz violation.
\newblock {\em Phys. Rev. Lett.}, 87:1416011--1416014, 2001.

\bibitem{Carosi90}
R.~{Carosi} et~al.
\newblock {A Measurement of the Phases of the CP-Violating Amplitudes in $K^0
  \rightarrow 2 \pi$ Decays and a Test of CPT Invariance}.
\newblock {\em Phys. Lett. B}, 237:303--312, 1990.

\bibitem{Schwingenheuer95}
B.~{Schwingenheuer} et~al.
\newblock {CPT Tests in the Neutral Kaon System}.
\newblock {\em Phys. Rev. Lett.}, 74:4376--4379, 1995.


\bibitem{Kostelecky98}
V.~Alan Kostelecky.
\newblock Sensitivity of CPT tests with neutral mesons.
\newblock {\em Phys. Rev. Lett.}, 80:1818--1821, 1998.

\bibitem{Kostelecky99}
V.~Alan Kostelecky.
\newblock Signals for CPT and Lorentz violation in neutral-meson oscillations.
\newblock {\em Phys. Rev.}, D61:0160021--0160029, 2000.

\bibitem{Kostelecky01b}
V.~Alan Kostelecky.
\newblock Formalism for CPT, T, and Lorentz violation in neutral- meson
  oscillations.
\newblock {\em Phys. Rev.}, D64:07600101--07600111, 2001.

\bibitem{Skadhauge02}
S.~{Skadhauge}.
\newblock Probing CPT violation with atmospheric neutrinos.
\newblock {\em Nucl. Phys.}, B639:281--289, 2002.

\bibitem{Barenboim02a}
G.~{Barenboim}, J.~F. {Beacom}, L.~{Borissov}, and B.~{Kayser}.
\newblock CPT violation and the nature of neutrinos.
\newblock {\em Phys. Lett.}, B537:227--232, 2002.

\bibitem{Barenboim02b}
G.~{Barenboim} and J.~{Lykken}.
\newblock A model of CPT violation for neutrinos.
\newblock {\em Phys. Lett.}, B554:73-80, 2003.

\bibitem{Barmin84}
V.~V. {Barmin} et~al.
\newblock {CPT Symmetry and Neutral Kaons}.
\newblock {\em Nuc. Phys. B}, 247:293--312, 1984; 
{\em  Erratum-ibid.} B254:747,1985 

\bibitem{Dejardin00}
M.~{D\'{e}jardin}.
\newblock T and CPT violation in the neutral-kaon system.
\newblock {\em Nuc. Phys. A}, 663:915--918, 2000.

\bibitem{Klinkhamer01}
F.~R. {Klinkhamer}.
\newblock Mechanism for CPT violation.
\newblock Prepared for International Europhysics Conference on High- Energy
  Physics (HEP 2001), Budapest, Hungary, 12-18 Jul 2001.

\bibitem{Urbanowski02}
K.~{Urbanowski}.
\newblock A new interpretation of one CPT violation test for K0 - anti-K0
  system.
\newblock {\em hep-ph/0202253}, 2002.

\bibitem{CERN02a}
M.~{Amoretti} et~al., ATHENA experiment,
\newblock  Production and detection of cold antihydrogen atoms.
\newblock {\em  Nature} 419:456--459, 2002.

\bibitem{CERN02b}
G.~Gabrielse  et~al., ATRAP experiment, 
\newblock Background-Free Observation of Cold Antihydrogen 
with Field-Ionization Analysis of Its States.
\newblock {\em Phys. Rev. Lett.},  89:2134011--2134014, 2002.

\bibitem{Letessier02}
J.~{Letessier} and J.~{Rafelski}.
\newblock {\em Hadrons and Quark-Gluon Plasma}.
\newblock Cambridge, 2002.

\end{thebibliography}

\end{document}